\documentstyle[psfig,conf_iap,]{article}
\begin{document}

\heading{%
Casimir effect and vacuum energy} 
\par\medskip\noindent

\author{%
Cyriaque Genet$^{1}$, Astrid Lambrecht$^{1}$ and Serge Reynaud$^{1}$ }

\address{%
Laboratoire Kastler Brossel UPMC/ENS/CNRS case 74, 
Campus Jussieu, F75252 Paris Cedex 05 }

\begin{abstract}
Vacuum fluctuations have observable consequences, like the Casimir force
appearing between two mirrors in vacuum. This force is now measured with
good accuracy and agreement with theory. We discuss the meaning and 
consequences of these statements by emphasizing their relation with the 
problem of vacuum energy, one of the main unsolved problems at the interface 
between gravitational and quantum theory.
\end{abstract}

\medskip The emergence of quantum theory has profoundly altered our
conception of empty space by forcing us to consider vacuum as permanently
filled by quantum field fluctuations. These fluctuations have numerous
observable effects in microscopic physics. They also have manifestations in
the macroscopic world, such as the Casimir force appearing between two
mirrors in vacuum as a consequence of the radiation pressure of vacuum
fluctuations. This force has recently been measured with a good accuracy and
in a correct agreement with theory, the latter taking into account the real
optical properties of the mirrors used in the experiments.

Before discussing the precise meaning of these statements, we will present
historical remarks devoted to the birth of the theory of vacuum fluctuations
as well as the difficult relation of their energy with gravitation theory.
We will also emphasize some features of the Casimir effect which
may be of interest in the cosmological context discussed in the present
volume \cite{IAP02}. The Casimir force is equivalent to a negative pressure
arising from the interaction of vacuum fluctuations with mirrors. Whereas
the vacuum energy density is infinite in the absence of cutoff and badly
defined (i.e. cutoff-dependent) otherwise, the Casimir pressure is well
defined and cutoff-independent. It is found to have a `weak' non vanishing
value, i.e. a value much smaller than any estimate of the `large' vacuum
energy density. Furthermore, this value agrees with the results of
experimental measurements.
We will finally discuss briefly other mechanical effects of vacuum
fluctuations which appear for moving mirrors. These effects also correspond
to perfectly well defined theoretical predictions which are in principle
observable, although their very small magnitude has up to now prevented
their experimental observation.

We begin with a few historical remarks related to the early history of
vacuum fluctuations. The classical idealization of space as being absolutely
empty was already affected by the advent of statistical mechanics, when it
was realized that space is in fact filled with black body radiation. It is
precisely for explaining the properties of this thermal radiation that
Planck introduced his first quantum law in 1900 \cite{Planck00}. In modern
terms, this law gives the mean energy per electromagnetic mode as the
product of the energy of a photon $\hbar \omega \equiv h\nu $ by a mean
number of photons $\overline{n}$ per mode 
\begin{equation}
\overline{E}_{\rm bbr}=\overline{n}\hbar \omega \qquad \overline{n}=
\frac{1}{\exp \left( \frac{\hbar \omega }{k_{\rm B}T}\right) -1}
\end{equation}
Here `bbr' labels the effect of black body radiation. Planck was certainly aware
of the unsatisfactory character of his first derivation. Like Einstein and
other physicists, he attempted for years to give a more satisfactory proof
of this law by studying in more detail the interaction between matter and
radiation. These attempts and the related discussions which led finally to the 
discovery by Einstein of the quantum absorption-emission laws and of the Bose 
statistics of photons \cite{Wolf79} are described for example in \cite{Milonni91}.

In a paper written in 1911, Planck \cite{Planck12} wrote a different
expression where the mean energy per mode $\overline{E}$ contains a term 
$\frac{\hbar \omega }{2}$ which describes a zero-point energy superimposed to
the black-body radiation energy 
\begin{equation}
\overline{E}=\overline{E}_{\rm bbr}+\frac{\hbar \omega }{2}  \label{eqEbar}
\end{equation}
In contrast with the thermal fluctuations which disappear at the limit of
zero temperature and can thus be\ `pumped out' of a cavity by lowering the
temperature, the zero-point fluctuations are still present at zero
temperature. The arguments used by Planck for deriving his second law can no
longer be considered as consistent today. Einstein was working along the same lines
at the same time and he noticed in a paper \cite{Einstein13} written with
Stern in 1913 that the first Planck law does not reproduce exactly the
classical limit at high temperature whereas the second law passes 
successfully this consistency check 
\begin{equation}
T\rightarrow \infty \qquad \overline{E}_{\rm bbr}\simeq k_{\rm B}T-
\frac{\hbar \omega }{2}+\frac{\hbar ^{2}\omega ^{2}}{12k_{B}T}+\ldots 
\qquad \overline{E}=k_{\rm B}T+O\left( \frac{1}{T}\right)
\end{equation}
This may considered as the first known argument still acceptable today: it
fixes the numerical factor $\frac{1}{2}$ in front of $\hbar \omega $, a
factor which was varying between 0 and 1 in the papers published at this
time. Amazingly, this argument fixes the magnitude of zero-point
fluctuations, essentially visible at low temperatures, by demanding their
disappearance in the high temperature limit to be as perfect as possible~!

It is worth emphasizing that physicists were numerous to take zero-point
fluctuations seriously, long before the latter were rigorously deduced from 
the fully developed quantum theory. As Planck, most of them
restricted their discussion of zero-point fluctuations to the case of atomic
motion. Nernst is credited for having been the first to emphasize that the
zero-point fluctuations should also exist for field modes \cite{Nernst16}.
In this sense, he may be considered as the father of `vacuum fluctuations',
that is precisely the zero-point fluctuations of the electromagnetic field.
He remarked that the very existence of these fluctuations dismisses the
classical idea of an empty space which may be attained, at least in
principle, by removing all matter from an enclosure and lowering the
temperature down to zero. He also noticed that this existence constitutes a
challenge for gravitation theory. When the vacuum energy density is
calculated by adding the energies $\frac{\hbar \omega }{2}$ over all field
modes, an infinite value is obtained. When a high frequency cutoff is
introduced, the sum is finite but still much larger than the mean energy
observed in the world around us through gravitational phenomena.

This major problem, which has been known since 1916, can be named the
`vacuum catastrophe' \cite{Adler95}, in analogy with the `ultraviolet
catastrophe' solved by Planck in 1900. The analogy is more than formal as
shown by the expression of the mean energy density calculated by summing 
(\ref{eqEbar}) over all field modes up to a high frequency cutoff $\omega _
{\rm max}$ 
\begin{equation}
\overline{\rho }=\overline{\rho }_{\rm bbr}+\overline{\rho }_{\rm vac}
\qquad \overline{\rho }_{\rm bbr}=\frac{\pi ^{2}\left( k_{\rm B}
T\right) ^{4}}{15\left( \hbar c\right) ^{3}}\qquad \overline{\rho }_
{\rm vac}=\frac{\left( \hbar \omega _{\rm max}\right) ^{4}}{8\pi ^{2}\left(
\hbar c\right) ^{3}}  \label{eqEnergyDensity}
\end{equation}
$\overline{\rho }_{\rm bbr}$ represents the energy density of black body
radiation and it is proportional to the fourth power $T^{4}$ of temperature,
in consistency with the Stefan-Boltzmann law. The fact that it has a finite
value independent of the cutoff is a great success of Planck theory of black
body radiation. But this success is ruined by the appearance of the second
term~! The vacuum energy density $\overline{\rho }_{\rm vac}$ is
proportional to the fourth power $\omega _{\rm max}^{4}$ of the cutoff, it
diverges when $\omega _{\rm max}$ goes to infinity and it is in any case
enormously larger than the observed energy density of empty space, for any
value of the cutoff compatible with quantum field theory at the energies
where the latter is well tested. This is directly connected to the `cosmological
constant problem' \cite{Weinberg89} which has remained unsolved up to now
despite considerable efforts for proposing solutions (see the discussions in
the present volume \cite{IAP02}).

The existence of vacuum fluctuations was confirmed by the quantum theory of
electromagnetic field \cite{Dirac58}. We do not need the complete theory in
the present paper and we will present instead a simple representation of
vacuum fluctuations used in the domain of quantum optics \cite{Reynaud92}.
To this aim, we represent the field ${\cal E}$ in any of the involved modes,
with a frequency $\omega $, as a sum ${\cal E}_{1}\cos \omega t+{\cal E}%
_{2}\sin \omega t$ of two components proportional to $\cos \omega t$ and 
$\sin \omega t$ respectively. ${\cal E}_{1}$ and ${\cal E}_{2}$ are called
the quadrature components of the field mode and they are quite analogous to
the position and momentum of an harmonically bound particle. In particular,
the two quadrature components are conjugated observables obeying an
Heisenberg inequality $\Delta {\cal E}_{1}\Delta {\cal E}_{2}\geq {\cal E}%
_{0}^{2}$. This is the basic reason for the necessity of quantum
fluctuations of electromagnetic fields, ${\cal E}_{0}$ being a constant
measuring the amplitude of these quantum fluctuations. By definition, vacuum
is simply the state corresponding to the minimum energy. Since the field
energy is proportional to ${\cal E}_{1}^{2}+{\cal E}_{2}^{2}$, this leads to
null mean fields $\left\langle {\cal E}_{1}\right\rangle =\left\langle 
{\cal E}_{2}\right\rangle =0$ and to equal and minimal variances for the two
quadratures $\Delta {\cal E}_{1}^{2}=\Delta {\cal E}_{2}^{2}={\cal E}%
_{0}^{2} $. This corresponds to the vacuum energy $\frac{\hbar \omega }{2}$
of the second Planck law (\ref{eqEbar}).

It is worth emphasizing that vacuum fluctuations are real electromagnetic
fields propagating in space with the speed of light, as any free field.
Since the vacuum state corresponds to the minimal energy of field
fluctuations, it is impossible to use this energy to build up perpetual
motions violating the laws of thermodynamics. But electromagnetic vacuum
fluctuations have well known observable consequences in atomic physics \cite
{Cohen88} and, more generally, in quantum field theory \cite{Itzykson85}. An
atom interacting only with vacuum fields suffers spontaneous emission
processes induced by these fields. When fallen in its ground state, the atom
can no longer emit photons but its coupling to vacuum still results in
measurable effects like the Lamb shift of absorption frequencies. Two atoms
located at different locations in vacuum experience an attractive Van der
Waals force which plays an important role in physico-chemical processes.
Casimir was studying this effect when he discovered in 1948 that a force
arises between two mirrors placed in vacuum \cite{Casimir48}. This effect is
discussed below.

These fluctuations have also been thoroughly studied through their effect on
the properties of photon noise. Photon noise indeed reflects the quantum
field fluctuations entering an optical system and they can be studied,
controlled and, even in some cases, manipulated by using specifically
designed devices \cite{Reynaud92}. This can be used for improving the noise
control in high sensitivity measurement apparatuses developed for example
for detecting gravitational waves or testing the equivalence principle in
space experiments \cite{Reynaud01}. In this respect, the status of vacuum
fluctuations in modern quantum theory is as firmly established as that of
most quantum phenomena.

However, the problem of gravitation of the energy of vacuum fluctuations has
persisted since Nernst. This point was crudely stated by Pauli in his
textbook on Wave Mechanics \cite{Pauli33}~:

\begin{quote}
{\it `At this point it should be noted that it is more consistent here, in
contrast to the material oscillator, not to introduce a zero-point energy of 
}$\frac{1}{2}h\nu ${\it \ per degree of freedom. For, on the one hand, the
latter would give rise to an infinitely large energy per unit volume due to
the infinite number of degrees of freedom, on the other hand, it would be
principally unobservable since nor can it be emitted, absorbed or scattered
and hence, cannot be contained within walls and, as is evident from
experience, neither does it produce any gravitational field'}
\end{quote}

A part of these statements is simply unescapable~: the mean value of vacuum
energy does not contribute to gravitation as an ordinary energy. This is
just a matter of evidence since the universe would look very differently
otherwise. However, it is certainly not possible to deduce, as Pauli did,
that vacuum fluctuations have no observable effects. Certainly, vacuum
fluctuations can `be emitted, absorbed, scattered...' as shown by their
numerous microscopic effects. When `contained within walls', they lead to
the observable, and in fact observed, Casimir effect discussed below. And
even if the reference level setting the zero of energy for gravitation
theory turns out to be finely tuned in the vicinity of the mean value of
vacuum energy density, energy differences and energy fluctuations still have
to contribute to gravitation, a point which will be discussed in the end of
this paper.

We now come back to the discussion of the Casimir force.
Casimir calculated this force in a geometrical configuration where two plane
mirrors are placed a distance $L$ apart from each other, parallel to each
other, the area $A$ of the mirrors being much larger than the squared
distance $\left( A\gg L^{2}\right) $. Casimir considered the ideal case of
perfectly reflecting mirrors and obtained the following expressions for the
force $F_{\rm Cas}$ and energy $E_{\rm Cas}$ 
\begin{equation}
F_{\rm Cas}=\frac{\hbar c\pi ^{2}A}{240L^{4}}\qquad E_{\rm Cas}=
\frac{\hbar c\pi ^{2}A}{720L^{3}}  \label{eqCasimir}
\end{equation}
The signs correspond to a convention opposite to the standard convention of
thermodynamics (but common in papers on the Casimir force)~: the force is
attractive and corresponds to a negative pressure; meanwhile, the
energy is a binding energy corresponding to a mean energy density slightly
smaller inside the cavity than in the outside vacuum. Note that the energy 
density and pressure obey the equation of state of pure radiation. 

The accuracy of the first experiments was limited but it has been
greatly improved in recent measurements (see reviews in \cite
{Bordag01,Lambrecht02}). Several experiments reached an experimental precision
at the \% level by using an atomic force microscope (AFM) \cite{Harris00} or
micro-electromechanical systems (MEMS) \cite{Chan01}. Furthermore, the
measurements agree with theory also at the \% level provided that deviations
from the ideal situation considered by Casimir are properly accounted for.

Let us first discuss the effect of imperfect reflection of the metallic
mirrors used in the experiments. Note that the ideal Casimir formulas (\ref
{eqCasimir}) only depend on the geometrical quantities $A$ and $L$ and on
two fundamental constants, the speed of light $c$ and Planck constant $\hbar 
$. This is a remarkably universal feature in particular because these
formulas are independent of the atomic constants, for instance the electron
charge $e$. This means that the Casimir force corresponds to a saturated
response of the mirrors which reflect 100 \% of the incoming light in the
ideal case, whatever their atomic constitution may be. But experiments are
performed with metallic mirrors which do not reflect perfectly all field
frequencies and this has to be taken into account in theoretical estimations 
\cite{Lambrecht00}. It follows that the force between real mirrors depends
on the properties of the latter.

Imperfectly reflecting mirrors are described by scattering amplitudes which
depend on the frequency, wavevector and polarization while obeying general
properties of stability, high-frequency transparency and causality. The two
mirrors form a Fabry-Perot cavity with the consequences well-known in
classical or quantum optics~: the energy density of the intracavity field is
increased for the resonant frequency components whereas it is decreased for
the non resonant ones. The Casimir force is but the result of the balance
between the radiation pressure of the resonant and non resonant modes which
push the mirrors respectively towards the outer and inner sides of the
cavity \cite{Jaekel91}. This balance includes not only the contributions of
ordinary waves propagating freely outside the cavity but also that of
evanescent waves. Using analyticity properties of the scattering amplitudes,
the Casimir force is finally written as an integral over imaginary
frequencies $\omega = i \xi$ (with $\xi$ real)
\begin{equation}
F=\frac{\hbar A}{\pi }\sum_{p}\int \frac{{\rm d}^{2}{\bf k}}{4\pi ^{2}} 
\int\limits_{0}^{\infty }{\rm d}\xi \frac{\kappa \ r_{1}r_{2}}{\exp \left(
2\kappa L\right) -r_{1}r_{2}}\qquad \kappa =\sqrt{{\bf k}^{2}+
\frac{\xi ^{2}}{c^{2}}}  \label{eqCasimirReal}
\end{equation}
$r_{1}$ and $r_{2}$ are the reflection amplitudes of the two mirrors as they
are seen from the intracavity fields and they are evaluated at imaginary
frequencies, for a transverse wavevector ${\bf k}$ and a
polarization $p$; $\kappa $ is the analytical prolongation of the
longitudinal wavevector which determines the dephasing corresponding to one
round trip. 

This expression holds for dissipative mirrors as well as non dissipative ones 
\cite{Genet02}. Thanks to high-frequency transparency, it is regular for any 
physical amplitudes $r_{1}$ and $r_{2}$, in spite of the infiniteness of vacuum 
energy. It goes to the Casimir formula (\ref{eqCasimir}) at the limit
where mirrors may be considered as perfect $\left( r_{1}r_{2}\rightarrow
1\right) $ over the frequency range of interest. Since metals are perfect
reflectors at frequencies lower than their plasma frequency, the real force 
(\ref{eqCasimirReal}) deviates from the ideal expression (\ref{eqCasimir}) at
distances $L$ shorter than a few plasma wavelengthes, typically $0.3\mu$m 
for gold or copper (for a more detailed discussion, see \cite{Lambrecht00}).

A second important correction is due to the geometry. The ideal formula 
(\ref{eqCasimir}) corresponds to the geometry of two parallel plane plates
whereas recent experiments have been performed with a sphere and a plane.
The force in the latter geometry is usually estimated from the proximity
theorem which basically amounts to integrate the force contributions of the
various inter-plate distances as if they were independent. The force
evaluated in this manner is thus given by the Casimir energy evaluated in
the plane-plane geometry for the distance $L$ of closest approach. Although
the accuracy of this approximation remains to be mastered, it is thought to
give a reliable approximation in the recent experiments \cite{Bordag01}. The
two already discussed corrections, respectively associated with imperfect
reflection and plane-sphere geometry, have a significant impact on the value
of the Casimir force and, thus, may be considered to be tested in the 
comparison between experimental measurements and theoretical expectations.

There are still further corrections taken into account in theory-experiment
comparisons. The first one is the radiation pressure of thermal fluctuations 
which are superimposed to vacuum fluctuations as soon as the temperature 
differs from zero. Basically, this correction amounts to replace in 
(\ref{eqCasimirReal}) the vacuum energy by the total energy (\ref{eqEbar}) 
of vacuum and black body fluctuations. The number of thermal photons is appreciable 
at low frequencies $\hbar \omega <k_{\rm B}T$ so that the thermal correction from
the ideal formula is significant at large distances, typically\ above $3\mu$m
at room temperature (for a more detailed discussion, see \cite
{Genet00,Klimchitskaya02}). As a consequence, it is smaller than 1\% for the most 
precise measurements which correspond to distances below $0.5\mu$m. The same
conclusion is reached for the surface roughness correction, here because it
would play a significant role at distances shorter than those explored in
the experiments \cite{Bordag01}. Since these two last corrections are small
in the most precise recent measurements, the corresponding effects cannot be 
considered to have been tested experimentally.

We may conclude the discussion of the Casimir force by saying that the
recent experimental and theoretical work have made possible an accurate
comparison, say near the 1\% level, between measurements and expectations.
This is important for the reason discussed in the introduction. The Casimir
force is the most accessible experimental consequence of vacuum fluctuations
in the macroscopic world and it is crucial to test it with the greatest care
and accuracy. Note also that mastering the Casimir force is a key point in a
lot of very accurate force measurements at distances between nanometer and
millimeter and which are motivated by searches for new short range weak
forces predicted in theoretical unification models (see \cite
{Bordag01,Reynaud01} for reviews).

In the sequel of this paper, we will discuss dynamical generalizations of
the Casimir effect. To be precise, the Casimir force
is a mechanical consequence of vacuum radiation pressure observed in the
presence of two static mirrors. There also exist such consequences for
mirrors moving in vacuum and they are directly connected to the
problem of relativity of motion. Even a single mirror isolated in vacuum 
experiences the fluctuations of vacuum radiation pressure
\cite{Fulling76,Ford82,Jaekel92QO,Barton94,MaiaNeto96}. When the mirror is
motionless, the resulting force has a null mean value due to the balance
between the contributions of opposite sides. When the mirror is moving, the
balance is broken for the mean force which means that vacuum exerts a
radiation reaction against motion. This reaction corresponds to a damping of
the mechanical energy which is associated with an emission of radiation by
the moving mirror into vacuum. The dissipative force is described by a
susceptibility allowing one to express the force $F(t)$ in a linear approximation 
versus the time-dependent position $q(t)$. This expression is more conveniently
written in the Fourier domain with $\Omega $ the frequency of mechanical motion 
\begin{equation}
F[\Omega ]=\chi [\Omega ] q[\Omega ]
\end{equation}
The motional susceptibility $\chi \lbrack \Omega ]$ is directly related to
the force fluctuations evaluated for a mirror at rest through quantum
fluctuation-dissipation relations \cite{Jaekel97}. We emphasize that it
describes dissipative effects, i.e. the radiation reaction force and
associated radiation emission, for a mirror moving without further reference
than the vacuum fluctuations \cite{Jaekel98}.

For comparison, let us consider a mirror moving through a thermal equilibrium 
field. It is well-known that a friction force proportional to the velocity 
$q^ \prime$ arises in this case. In the frequency domain, the susceptibility is 
thus proportional to the frequency~: for a perfectly reflecting mirror in the
limits of large plane area and large temperature $\left( A\gg \frac{c^{2}}
{\Omega ^{2}}\ ,\ k_{\rm B}T\gg \hbar \Omega \right) $, it is read as
\begin{equation}
\chi _{\rm bbr}[\Omega ] = \frac{i\pi ^{2}\hbar A}{15}\left( \frac{k_
{\rm B}T}{\hbar c}\right) ^{4}\Omega \qquad F_{\rm bbr}(t) = - \frac{\pi
^{2}\hbar A}{15}\left( \frac{k_{\rm B}T}{\hbar c}\right) ^{4} q^ \prime (t)
\label{eqMotionalHighT}
\end{equation}
The damping coefficient is proportional to the fourth power of temperature,
again in full conformity with the Stefan-Boltzmann law, and it vanishes when 
$T$ goes to zero. This does not lead to the absence of any dissipative
effect of vacuum on a moving scatterer. For a perfectly reflecting mirror in
the limits of large plane area and low temperature $\left( A\gg \frac{c^{2}}
{\Omega ^{2}}\ ,\ k_{\rm B}T\ll \hbar \Omega \right) $, we obtain 
\begin{equation}
\chi _{\rm vac}[\Omega ] = \frac{i\hbar A}{60\pi ^{2}c^{4}}\Omega
^{5}\qquad F_{\rm vac} (t) = -\frac{\hbar A}{60\pi ^{2}c^{4}}
q^{\prime \prime \prime \prime \prime }(t)  \label{eqMotionalZeroT}
\end{equation}
The susceptibility is now proportional to the fifth power of frequency which
means that the force is proportional to the fifth order time derivative of the
position. In particular, the radiation reaction force vanishes in the case
of uniform velocity, so that the reaction of vacuum cannot distinguish
between inertial motion and rest, in full consistency with the principle of
special relativity of motion. The quantum formalism gives an interesting
interpretation of this property~: vacuum fluctuations are preserved under
Lorentz transformations and they appear exactly the same to an inertial
observer as to an observer at rest. At the same time, the existence of
dissipative effects associated with arbitrary motion in vacuum challenges
the principle of relativity of motion in its more general acceptance. The
space in which motion takes place can no longer be considered as absolutely
empty since vacuum fluctuations are always present, giving rise to real
dissipative effects. In this sense, vacuum fluctuations allow us to define
privileged reference frames for the definition of mechanical motions. 

Note that the form of the last result can be guessed from the previous one
through a mere dimensional analysis~: the effect of temperature $T$ appears
in (\ref{eqMotionalHighT}) measured as the fourth power $\left( \frac{k_
{\rm B}T}{\hbar }\right) ^{4}$ of a frequency which is replaced by $\Omega
^{4}$ in (\ref{eqMotionalZeroT}). A similar dimensional argument appeared in
the discussion of the expression (\ref{eqEnergyDensity}) of vacuum energy
density with $\left( \frac{k_{\rm B}T}{\hbar }\right) ^{4}$ replaced by 
$\omega _{\rm max}^{4}$. There is however an important difference~: the
vacuum term which was ill-defined in (\ref{eqEnergyDensity}) has now a regular 
cutoff-independent expression while being associated with well-defined physical
effects. This is a further example, after the static Casimir effect, where
we obtain a `weak' (with respect to the `large' vacuum energy density) but
non vanishing value for an effect induced by vacuum fluctuations. 

Clearly, it would be extremely interesting to obtain experimental evidence
for the dissipative effects associated with motion in vacuum. These effects
are exceedingly small for any motion which could be achieved in practice for
a single mirror, but an experimental observation is conceivable with a
cavity oscillating in vacuum. In this case, the emission of motional
radiation is resonantly enhanced \cite{Lambrecht96} and specific signatures
are available for distinguishing the motional radiation from spurious
effects \cite{Lambrecht98}. Hence, an experimental demonstration appears to
be achievable with very high finesse cavities \cite{Jaekel01}.

The radiation reaction force associated with the motion of a cavity in
vacuum contains an inertial contribution in the specific case of uniformly
accelerated motion. This effect is a quantum version, at the level of vacuum
fluctuations, of Einstein argument for the inertia of a box containing a
photon bouncing back and forth. Here, Einstein law of inertia of energy has
to be applied to the case of a stressed body, where it is read \cite
{Einstein07} 
\begin{equation}
F_{\rm vac}=-\mu a\qquad \mu =\frac{E_{\rm Cas}-F_{\rm Cas}L}{c^{2}}
\end{equation}
The explicit calculation of the vacuum radiation reaction force gives an
expression which perfectly fits this law \cite{Jaekel93,Machado02}. This
means that the Casimir energy, which is a variation of the vacuum energy,
contributes to the inertia of the Fabry-Perot cavity as expected from the
general principles of relativity. Again this entails that, even if the mean
value of vacuum energy does not contribute to gravity, energy differences
such as the Casimir energy are expected to contribute to gravitational or
inertial phenomena \cite{Moffat02,Caldwell02}.

\acknowledgements{We thank G.~Barton, F.~Capasso, E.~Fischbach, M.-T.~Jaekel, 
S.~Lamoreaux, J.~Long, P.A.~Maia Neto, U.~Mohideen and R.~Onofrio for 
stimulating discussions.}

\vfill

\end{document}